%% file: make_astro.tex
\documentclass{mpe_report}

\usepackage{psfig,graphicx,epsfig}
\usepackage{color}
\usepackage{amsmath,amssymb,epic,eepic,array}

\begin{document}

\pagenumbering{arabic}
\setcounter{page}{141}

 \renewcommand{\FirstPageOfPaper }{141}\renewcommand{\LastPageOfPaper }{144}\include{./mpe_report_hsu}                  \clearpage

\end{document}

%% file: mpe_report_hsu.tex

\title{Particle Acceleration in Pulsar Magnetospheres}
\author{Pei-Chun Hsu\inst{1}, Kouichi Hirotani\inst{2} \and Hsiang Kuang Chang\inst{3}}
\institute{Institude of Astronomy, National Tsing Hua University, Taiwan, R.O.C.
\and Institute of Astronomy and Astrophysics, Academia Sinica, Taiwan, R.O.C.
\and Department of Physics \& Institude of Astronomy, National Tsing Hua University, Taiwan, R.O.C.}
\maketitle

\newcommand{\rlc}{\varpi_{\rm LC}} 
\newcommand{\Ell}{E_\parallel}      
\newcommand{\inc}{\alpha_{\rm i}}  
\newcommand{\rhoGJ}{\rho_{\rm GJ}}  

\begin{abstract}
We investigate a pair creation cascade in the magnetosphere of a rapidly rotating neutron star. 
We solve the set of the Poisson equation for the electro-static potential and the Boltzmann equations for electrons, positrons, and gamma-ray photons simultaneously. In this paper, we first examine the time-dependent nature of particle accelerators by solving the non-stationary Boltzmann equations on the two-dimensional poloidal plane in which both the rotational and magnetic axes reside. Evaluating the temperature of the heated polar cap surface, which is located near the magnetic pole, by the bombardment of gap-accelerated particles, and applying the scheme to millisecond pulsar parameters, we demonstrate that the solution converges to a stationary solution of which pair-creation cascade is maintained by the heated polar-cap emission, in a wide range of three-dimensional parameter space (period, period derivative, magnetic inclination angle). We also present the deathlines of millisecond pulsars.
\end{abstract}

\section{Introduction}

The Energetic Gamma-Ray Experiment Telescope (EGRET) on the Compton
 Gamma Ray Observatory (CGRO) had detected gamma-ray pulsed emissions
 from seven pulsars.  The light curves and spectra of these gamma-ray
 pulsars enable us to understand the $\gamma$-ray emission 
mechanism and examine different models. The prospective $\gamma$-ray 
telescope such as Gamma-ray Large Area Space Telescope (GLAST)
will provide more $\gamma$-ray observation with higher sensitivity. 
The origin of gamma-ray emission from pulsar magnetosphere will be 
studied more thoroughly.

Base on the work Cheng, Ho, \& Ruderman (1986a, b), the three-dimensional 
outer gap models (Romani \& Yadigaroglu 1995; Cheng, Ruderman and Zhang 2000)
can successfully explain the observed light curve in reproducing the wide 
separation of the two peaks with bridge emission.
Dyks and Rudak (2003) found that radiation region with inner boundary extended
 toward the stellar surface can explain the off-pulse emission, which
 cannot be reproduced by a traditional outer gap. 
An one-dimensional model 
of an outer gap with electrodynamical approach was investigated
(Hirotani \& Shibata 1999a, b, c; Hirotani, Harding \& Shibata 2003). 
They solved accelerating
 electric field, curvature radiation, and pair creation processes 
self-consistently. The model spectra can fit Vela observation well
at 100 MeV -- 10 GeV. In their model, it was analytically demonstrated that an 
active gap, which must be non-vacuum, possesses a qualitatively different 
properties from the vacuum solution of traditional outer-gap models. It was 
shown that the inner boundary of the outer gap shifts toward the stellar 
surface as the created current increases.

Subsequently, Takata, Shibata, and Hirotani (2004)
and Takata et al. (2006) studied the two-dimensional
 model which takes the trans-field structure into account. 
Solving the Poisson equation for the electrostatic potential
on the two-dimensional poloidal plane,
they revealed that the gap inner boundary is located inside of the
null surface because of the pair creation within the gap.
In their model, it is assumed that the particle motion immediately saturates
in the balance between electric and radiation-reaction forces.
However, there are cases in which particles run a good fraction of 
the whole gap length until they achieve the saturation. 
On these grounds, we solve the unsaturated
particle momenta and pitch angle evolution with synchrotron-curvature 
radiation in the two-dimensional geometry and we will discuss 
the particle motion and the electric field structure of the outer gap 
in this paper.

The high energy emission from millisecond pulsars (MSPs) is another 
important issue. 
Because of their short period, small period derivatives 
($\dot{P}<10^{-17}$s s$^{-1}$) as well as their evolutional history 
in the binary systems, MSPs form a distinctive group from normal pulsars. 
Thus, it is worth constructing a self-consistent electrodynamic model
for high-energy emission from MSPs.
Since the electrodynamic outer-gap models 
(Hirotani, Shibata \& Takata 2004 ; Takata et al. 2006 ; Hirotani 2006a) have applied the method only to
young and middle-aged pulsars (e.g., Crab, Vela, Geminga),
and used whole surface cooling neutron star thermal emission,
we investigate the electrodynamic model of MSPs in this paper,
evaluating the soft-photon field from the heated polar cap.

\section{Electrodynamics of Particle Accelerator}  
\subsection{Poisson equation}
The Poisson equation of the non-corotational potential $\phi$ is given by
\begin{equation}
\nabla^2\phi= -4\pi(\rho-\rho_{GJ})~,
\end{equation}
where the Goldreich-Julian charge density is defined as
(Goldreich \& Julian 1969)
\begin{equation}
  \rho_{\rm GJ}
  =~-~\frac{\vec{\Omega}\cdot \vec{B}}{2\pi c}
    +\frac{(\vec{\Omega}\times\vec{B})(\nabla\times\vec{B})}{4\pi c}~.
\end{equation}

\subsection{particle equations of motion}
\label{sec:EOMp}
The evolutions of momentum ($p\equiv|\vec{\mathstrut p}|$) and 
pitch angle ($\chi$) of a positron and an electron are described by
\begin{equation}
\frac{d}{dt}p=qE_{||}\cos\chi-\frac{P_{sc}}{c}
\label{y1}
\end{equation}
\begin{equation}
\frac{d}{dt}\chi=\frac{qE_{||}\sin\chi}{p}
\label{y2}
\end{equation}
\begin{equation}
\frac{d}{dt}s=\frac{p\cos\chi}{\sqrt{m_e^2+(p/c)^2}}~,
\label{y3}
\end{equation}
where $c$ is the speed of light. 
For positrons (or electrons), $q=+e$ (or $q=-e$) is adopted, 
where $e$ is the magnitude of the charge on an electron.
$P_{\rm SC}/c$ is the radiation reaction force 
due to synchro-curvature radiation. 

\subsection{Two-photon pair creation rate}
\label{sec:InitCond}
The electron-positron pairs are created via photon-photon pair creation. 
The creation rate $S_+(t, \vec{\mathstrut r}, \vec{\mathstrut p})$ 
for positrons and $S_-(t, \vec{\mathstrut r}, \vec{\mathstrut p})$ 
for electrons with momentum $\vec{\mathstrut p}$ 
at position $\vec{\mathstrut r}$ and time t is given by
\begin{eqnarray}
  \ S_\pm(t,\vec{\mathstrut r},\vec{\mathstrut p}) = \int_{-1}^{1} d\mu_{\rm c} \int d\epsilon_\gamma \int d\vec{\mathstrut k}~  \eta_{p\pm} 
             g(\epsilon_\gamma,\vec{\mathstrut r},\vec{\mathstrut k})~,
  \label{eq:def_S}
\end{eqnarray}
where the function $g$ represents the $\gamma$-ray distribution function 
at energy $m_e c^2 \epsilon_\gamma$, momentum $\vec{\mathstrut k}$, 
and position $\vec{\mathstrut r}$. 
The pair creation redistribution function $\eta_p$ can be defined as
\begin{eqnarray}
&&
\eta_{p\pm}(\epsilon_\gamma,\mu_{\rm c},\vec{\mathstrut p},\vec{\mathstrut k})
\nonumber\\
&&=
(1-\mu_c)\int_{\epsilon_{th}}^{\infty}d\epsilon_x\frac{dN_x}{d\epsilon_x}\sigma_p(\epsilon_\gamma,\epsilon_x,\mu_c)\delta(\vec{\mathstrut p}-\vec{\mathstrut k}/2)~,
\label{etap}
\end{eqnarray}
where $\mu_c$ is the cosine of the collision angle 
between the two photons,
$\epsilon_{\rm th} \equiv 2/[(1-\mu_{\rm c})\epsilon_\gamma]$ 
is the dimensionless threshold energy 
for a X-ray photon having energy $m_e c^2 \epsilon_x$
to make a pair with the $\gamma$-ray having 
energy $m_e c^2\epsilon_\gamma$;
$\sigma_p$ refers to the pair-creation cross-section.

\subsection{$\gamma$-ray emission, propagation, and absorption}
\label{sec:EOMg}
The probability for a charged particle (with charge $e$) to emit photons via pure-curvature process 
is given by (Rybicki \& Lightman 1979)
\begin{equation}
  \eta_c
  =\frac{\sqrt{3}e^2\Gamma}{hR_{\rm c}}\frac{1}{\epsilon_\gamma}
   F(\frac{\epsilon_\gamma}{\epsilon_c})~,
  \label{etac}
\end{equation}
\begin{equation}
  \epsilon_c=\frac{1}{m_ec^2}\frac{3}{4\pi}\frac{hc\Gamma^3}{Rc}~,
  \label{ec}
\end{equation}
\begin{equation}
  F(s)=s\int_{x}^{\infty}K_{5/3}(t)dt~,
  \label{f}
\end{equation}
where $\epsilon_c $ is the critical energy of curvature radiation,  
$R_{\rm c}$ the curvature radius of the magnetic field line, 
and $K_{5/3}$ the modified Bessel function of the order 5/3.

A $\gamma$-ray photon is emitted tangential to the local magnetic field line. 
The photon propagates along a straight ray path because we assume the flat 
space-time geometry, which is a good approximation in the outer magnetosphere. 

During the propagation, a $\gamma$-ray may collide with the thermal X-ray and 
be absorbed due to the pair creation process as described in the 
foregoing section. The absorption rate is given by
\begin{eqnarray} 
  S_\gamma
  &=& - \int_{-1}^{1} d\mu_{\rm c} 
        \int_1^\infty d\Gamma
           \eta_p(\vec{\mathstrut r},\Gamma,\mu_{\rm c})
           \cdot g(\epsilon_\gamma,\vec{\mathstrut r},
	           \vec{\mathstrut k})~.
\end{eqnarray} 

\subsection{boundary conditions}
\label{sec:BDC}
We consider the situation that the lower boundary of the gap, $\theta_\ast = \theta_\ast^{\rm max}$, coincides with the last open field line. 
$\theta_*$ is the magnetic colatitude of the point where the magnetic field line intersects the stellar surface.
We assume that the upper boundary, $\theta_\ast=\theta_\ast^{\rm min}$, coincides with a particular magnetic field line between the last open field line and the magnetic axis.The thickness of the gap is defined as
\begin{equation}
h_m \equiv \frac{\theta_\ast^{\rm max}- \theta_\ast^{\rm min}}{\theta_\ast^{\rm max}}~.
\end{equation}

We solve the Poisson equation from the neutron star surface to the
outer magnetosphere.
Thus, we define that the inner boundary coincides with the stellar surface.
As for the {\it outer} boundary,
we solve the Poisson equation to a large enough distance, 
$s=1.4~\rlc$, which is located outside of the light cylinder.
Since the structure of the outer-most part of the magnetosphere 
is highly unknown, 
we artificially set $\Ell=0$ if the distance from the rotation axis, 
$\varpi$, becomes greater than $0.90~\rlc$.
Under this artificially suppressed $\Ell$ distribution in $\varpi>0.90~\rlc$,
we solve the Boltzmann equations for outward-migrating 
particles and $\gamma$-rays in $0<s<1.4~\rlc$.
For inward-migrating particles and $\gamma$-rays, we solve only in
$\varpi<0.9~\rlc$.

To solve the Poisson equation, 
an elliptic-type partial differential equation,
we impose $\phi=0$ at the inner, lower, and upper boundaries, 
assuming that both the lower, and upper boundaries are grounded to the star.  
At the outer boundary, we impose $\partial\phi/\partial s = -E_\parallel =0$.

\section{Result}
\subsection{Time-dependent Simulation}
It is, unfortunately, difficult to constrain the time-dependent feature of 
$\gamma$-ray emissions observationally, 
because the EGRET experiment detected 
one GeV photon every tens of thousand pulsar rotations.
Taking advantage of this fact, 
all the high-energy emission theories from rotating neutron-star 
magnetospheres have concentrated on stationary analyses.
This drives us to the question whether the particle accelerators are,
indeed, stationary or not, which is still unsettled.

On these grounds, 
we solve the Poisson equation and time-dependent Boltzmann equations for 
electrons, positrons and gamma-ray photons self-consistently. 
By starting from a nearly vacuum solution,
we find that the solution evolves to a stationary state 
in a wide range of pulsar parameters,
such as period, $P$, period derivative $\dot{P}$, and
magnetic inclination angle, $\inc$.
We will demonstrate this new result, applying the method to 
millisecond pulsar parameters in the subsections below.

\subsection{Stability of Outer-Magnetospheric Gaps}

It is conjectured by Hirotani (2006) that an outer-magnetospheric gap is electrodynamically stable. In this subsection, we will examine this prediction by a time-dependent simulation. We start with the vacuum solution of $E_{||}(s,\theta_\ast)$ with an injection of electrons at 1\% of Goldreich-Julian rate, 0.01 $\Omega B/2\pi$ [cm$^{-2}$s$^{-1}$], across the outer boundary.

Such injected, inward-accelerated electrons emit copious curvature $\gamma$-rays, which efficiently collide with the surface X-rays to materialize as pairs in the inner magnetosphere. The created pairs discharge, partially screening the original $E_{||}$. The discharged positrons are accelerated outwards, emitting curvature photons, some of which materialize in the outer magnetosphere. The discharged electrons, on the other hand, emit inward-directed $\gamma$-rays, which efficiently cascade into higher generation pairs by head-on collisions with the surface X-rays. 

If the created pair density becomes too large, the over-screened $E_{||}$ results in a less-efficient $\gamma$-ray emission, leading to a decrease of pairs in a dynamical time scale, $\varpi_{\rm LC}/c$. On the other hand, if the created pair density is too small, the under-screened $E_{||}$ results in an increase of pairs. On these grounds, we can expect that an initial trial function evolves to the final, stationary solution after a lapse of several dynamical time scales.

\begin{figure}[h]
 \centerline{\psfig{file=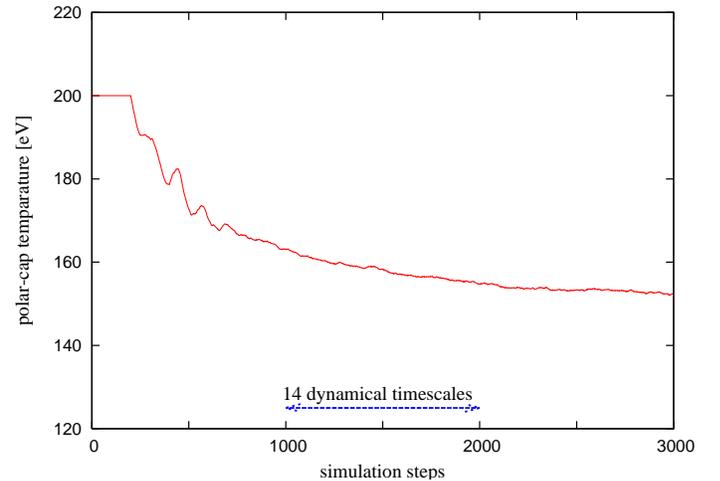,width=6.5cm,angle=-90,clip=} }
    \caption{The heated polar-cap temperature evolution. }
    \label{FigTemp}
\end{figure}

In figure \ref{FigTemp}, we present the result of a stabilized solution: The heated polar-cap temperature settles down to the final value $kT\sim 153$ eV, after about 20 dynamical timescales. We adopted $P=3.3$ ms, $\dot{P}=6.2\times 10^{-20}~{\rm s\,s}^{-1}$, and $\alpha_i=75^\circ$, with $h_m=0.9$.

We also find that this solution is marginally self-sustained in the sense that 
if one of the parameter is changed such as $P>3.3$ ms, or $\dot{P}<6.2\times 10^{-20}~{\rm s\,s}^{-1}$, or $\alpha_i<75^\circ$, or $h_m<0.9$, 
the created current density evolves to a negligibly small value compared to the Goldreich-Julian value; 
in this case, the surface temperature is simply attributed to the injected particles across the outer boundary and kept below $\sim50$ eV.

On the other hand, for a faster spin (i.e., $P<3.3$ ms), or larger magnetic moment (i.e., $\dot{P}>6.2\times10^{-20}~{\rm s\,s}^{-1}$), the final stable solution, with created current comparable to the Goldreich-Julian current and significantly greater than the injected current, has a higher polar-cap temperature, $kT>153$ eV.

\subsection{Millisecond pulsar death line}
Let us finally enlarge the electrodynamic consideration into the criterion for a gap to be self-sustained.
If particles are efficiently accelerated and the resultant created current density, $j_{\rm gap}$, is large enough, the electrons flowing onto the polar-cap surface will lead to a sufficient thermal emission from the surface. The efficient thermal photon flux ensures copious pair creation in the gap, and hence the large enough $j_{\rm gap}$ self-consistently. 

However, if the external electric field, $E_{||}$, decreases by some fictitious mechanism 
(e.g., by artificial decrease of neutron-star magnetic moment, $\mu$), $j_{\rm gap}$ and hence the polar-cap temperature, $kT$, decrease. The decrease of the surface photon flux further decreases $j_{\rm gap}$, which partially recovers the decrease of $E_{||}$. However, if $E_{||}$ is decreased (e.g., by the decrease of $\mu$) too much, even the vacuum solution of $E_{||}$ cannot maintain the pair creation cascade in the gap; in this case, $j_{\rm gap}$ as well as $kT$ tends to vanish, remaining an in-active gap with the vacuum potential drop.

Viewed in this light, we can define a 'death line' of the particle accelerator on $(P, \dot{P})$ plane, adopting some fixed value of $\alpha_i$ and $h_{\rm m}$. If pair creation is insufficient, then the gap upper boundary will be located near the magnetic pole, because the upper boundary will be formed by the copiously created charges that freely moves along the field lines and screens $E_{||}$. Thus, to find the death line, we adopt a large value of $h_m=0.9$ in this subsection.
As for $\alpha_i$, we adopt two representative values, $60^\circ$ and $75^\circ$. 

\begin{figure}[h]
\centerline{\psfig{file=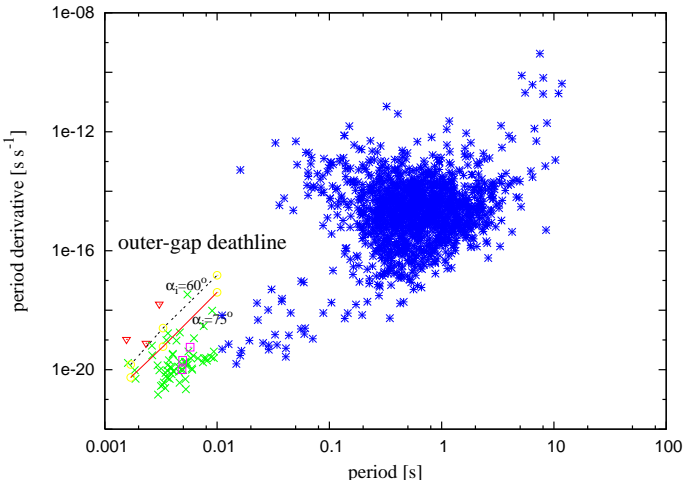,width=6.5cm,angle=-90,clip=} }
    \caption{$P-\dot{P}$ diagram of pulsars and millisecond deathline. Pulsars with period shorter than 10 ms are denoted by crosses. The dashed and solid lines denote the deathline for inclination angle 60 and 75 degree. The triangles denote the Class-II millisecond pulsars while the squares denote Class-I.
}
    \label{FigDeathline}
\end{figure}

To determine the death line, we investigate the marginal value of $\dot{P}$ below which a self-sustained gap does not exist for a fixed value of $P$. For both cases, $\alpha_i=60^\circ$ and $75^\circ$, we adopt three fixed values of $P$, 1.7 ms, 3.3 ms, 10 ms. For $\alpha_i$=$60^\circ$, we find that $\dot{P}>1.5\times 10^{-20}$ for $P=1.7$ ms, $\dot{P}>2.5\times 10^{-19}$ for $P=3.3$ ms, and $\dot{P}>1.5\times 10^{-17}$ for $P=10$ ms. For $\alpha_i=75^\circ$, on the other hand, $\dot{P}>6.0\times 10^{-21}$ for $P=1.7$ ms, $\dot{P}>6.0\times 10^{-20}$ for $P=3.3$ ms, and $\dot{P}>4.0\times 10^{-18}$ for $P=10$ ms. These conditions are plotted on $(P,\dot{P})$ diagram in figure \ref{FigDeathline} with open circles. The inferred death lines from these six points are also shown in figure \ref{FigDeathline}: the dashed line denotes the death line for $\alpha_i=60^\circ$ while the solid line for $\alpha_i=75^\circ$. 

It follows from figure \ref{FigDeathline} that the three Class II X-ray MSPs, denoted by triangles, appear above the two death lines. On the contrary, Class I  X-ray MSPs (squares) and most of other MSPs (crosses) are located below the death lines. 
Since the characteristics of Class-II members point to a non-thermal origin due to active particle acceleration taking place in the magnetosphere of a MSP, it is natural to interpret that the current calculation is associated with the Class-II MSPs. 
It would be possible to argue that the consistency of the obtained death line with the location of Class II MSPs on $(P,\dot{P})$ diagram support this interpretation of their non-thermal emission as a result of the self-sustained magnetospheric particle accelerator investigated in the present paper.

\begin{acknowledgements}
This work was supported by the National Science Council of the Republic of
China with the
grant NSC 94-2112-M-007-002.
\end{acknowledgements} 

\newcommand{\apj}{\rm ApJ}
\newcommand{\apjl}{\rm ApJL}
\newcommand{\mnras}{\rm MNRAS}
